\documentclass[aps,pra,twocolumn,superscriptaddress,showpacs]{revtex4-1}  % for review and submission
\usepackage{graphicx}  % needed for figures
\usepackage{epsfig}
\usepackage{dcolumn}   % needed for some tables
\usepackage{bm}        % for math
\usepackage{amssymb,amsmath,amsopn,amsfonts}
\usepackage{colordvi}
\usepackage{color}
\usepackage{multirow}
\usepackage{subeqnarray}
\usepackage{cases}
\newcommand{\ket}[1]{\ensuremath{\left| #1 \right\rangle}}

\def\eqref#1{\textup{(\ref{#1})}}  %% overiding the original command \eqref
\newcommand{\eref}[1]{Eq.~\textup{(\ref{#1})}}

\newcommand{\esref}[1]{Eqs.~\textup{(\ref{#1})}}

\newcommand{\vect}[1]{\boldsymbol{#1}}

\begin{document}

\title{Error-compensation measurements on polarization qubits}
\author{Zhibo Hou}
\affiliation{Key Laboratory of Quantum Information,University of Science and Technology of China, CAS, Hefei 230026, P. R. China}
\affiliation{Synergetic Innovation Center of Quantum Information and Quantum Physics, University of Science and Technology of China, Hefei 230026, P. R. China}
\author{Huangjun Zhu}
\email{hzhu@pitp.ca}
\affiliation{Perimeter Institute for Theoretical Physics, Waterloo, On N2L 2Y5, Canada}
\author{Guo-Yong Xiang}
\email{gyxiang@ustc.edu.cn}
\affiliation{Key Laboratory of Quantum Information,University of Science and Technology of China, CAS, Hefei 230026, P. R. China}
\affiliation{Synergetic Innovation Center of Quantum Information and Quantum Physics, University of Science and Technology of China, Hefei 230026, P. R. China}
\author{Chuan-Feng Li}
\affiliation{Key Laboratory of Quantum Information,University of Science and Technology of China, CAS, Hefei 230026, P. R. China}
\affiliation{Synergetic Innovation Center of Quantum Information and Quantum Physics, University of Science and Technology of China, Hefei 230026, P. R. China}
\author{Guang-Can Guo}
\affiliation{Key Laboratory of Quantum Information,University of Science and Technology of China, CAS, Hefei 230026, P. R. China}
\affiliation{Synergetic Innovation Center of Quantum Information and Quantum Physics, University of Science and Technology of China, Hefei 230026, P. R. China}

\date{\today}

\begin{abstract}
Systematic errors are inevitable in most measurements performed in real life because of imperfect measurement devices.
    Reducing systematic errors is crucial to ensuring the accuracy and reliability of  measurement results. To this end, delicate  error-compensation design is often necessary in addition to device calibration to reduce the dependence of the systematic error on the imperfection of the devices.  The art of error-compensation design is well appreciated in nuclear magnetic resonance  system by using composite pulses. In  contrast, there are few works on reducing systematic errors in quantum optical systems.
     Here we propose an  error-compensation design  to reduce the systematic error in  projective measurements on a polarization qubit.
     It can reduce the systematic error
     to the second order of the phase errors of both the half-wave plate (HWP) and the quarter-wave plate (QWP) as well as the angle error of the HWP. This technique is then applied to experiments on quantum state tomography on polarization qubits, leading to a 20-fold  reduction in the  systematic error. Our study may find applications in high-precision tasks in polarization optics and quantum optics.

\end{abstract}


%\pacs{03.67.-a, 03.65.Wj, 06.20.Dk, 03.65.Ta}
%03.67.-a: quantum information
%03.65.Wj: quantum tomography, state reconstruction
%03.65.-w: quantum mechanics
%03.65.Ta: Foundations of quantum mechanics

%06.20.Dk: Measurement and error theory
%42.50.Dv: Quantum state engineering and measurements
%42.25.Ja Polarization

%03.67.Mn: Entanglement production, characterization and manipulation
%03.65.Ud: Entanglement and quantum nonlocality
%(e.g. EPR paradox, Bell's inequalities, GHZ states, etc.)
%(for entanglement production in quantum information, see 03.67.Mn;

%02.10.De: algebraic structure
\maketitle

\section{Introduction}

Quantum measurement \cite{Saku85modern} is a cornerstone of quantum mechanics. It is also a basic ingredient of many
quantum information processing protocols, such as quantum computation \cite{Niel00quantum}, quantum communication \cite{Gisi02quantum}, and quantum metrology \cite{Giov04quantum,Giov06quantum,Giov11advances,Higg07entanglement}. Accurate implementation of quantum measurements is crucial  to tasks that  demand high precision, such as quantum state tomography \cite{Pari04quantum,Jame01measurement,Lvov09continuous,Krav13experimental,Mahl13adaptive,Zhu12quantum} and quantum process tomography \cite{Anis12maximum,Alte03ancilla,Kim14quantum,Brie04quantum}. In practice, however, measurement devices are never perfect, and the accuracy of measurements is often limited  by systematic errors. Detection and correction of systematic errors are thus crucial to ensuring the accuracy and reliability of measurement results.

There are already many works on the detection  and certification of systematic errors \cite{Schw11detecting,lang13errors,Mogi13cross,Van13quantum, Kimm14robust, Moro13certifying,Yang14robust,Moha14optimization}.
However, few works have addressed the problem of
reducing systematic errors induced by  imperfect devices. There are mainly two approaches for dealing with this problem. The simplest and most direct way is to reduce the imperfection of the devices by calibration. However, accurate calibration requires  accurate calibration apparatuses, which are not always available. In addition,  calibration is difficult  to conduct on large systems. The other approach  works by alleviating the dependence of the systematic error on the imperfection of the devices rather than improving the quality of the devices. This is usually achieved  via an elaborate design to correct the systematic error to the second or even a higher order of the device imperfections.

In nuclear magnetic resonance (NMR) system, many delicate composite pulse sequences have been designed  to compensate a variety of systematic errors \cite{Levi79nmr,Tyck83broadband}. Certain composite pulse sequences can  correct the systematic errors to an arbitrary order \cite{Brow04arbitrarily,Low14optimal} and apply to any initial state \cite{Levi86composite}. Similar method has also been used to combat decoherence noise \cite{Cumm00use,Guld03implementation,Coll04nmr,Roy12dynamical,Roy13dynamical}.
By contrast, little is known about compensating systematic errors in quantum optical devices. One exception is the work of Ardavan \cite{Arda07exploiting} (see also Refs.~\cite{Ivan12highly,Pete12variable,Rang15broadband}), in which high-fidelity broadband composite linear retarders were designed following the idea of composite pulses in NMR system to combat chromatic dispersion.  This approach assumes that  phase errors of all the linear retarders are the same and only applies to errors induced by chromatic dispersion. Incidentally, another approach for making wave plates less sensitive to the  wavelength is to combine two wave plates made of materials with different birefringences \cite{Clar67achromatic}.
In practice,  linear retarders are not perfect, and their phase errors may differ from each other. For example, typical phase errors of current phase retarders on the market are around $\lambda/300$; they   dominate over chromatic dispersion if the wavelength bandwidth is less than $\lambda/150$, which is usually the case. Unfortunately, the approaches mentioned above  cannot correct such phase errors; neither can they tackle with angle errors resulting from imperfect calibration of optic axes.

In this paper, we present a simple method for correcting systematic errors in projective measurements  on polarization qubits. The measurement device we consider is composed of  a quarter-wave plate (QWP), a half-wave plate (HWP), and a polarizing beam splitter (PBS), referred to  as the QWP-HWP setting henceforth. Our strategy consists in designing a series of  rotation-angle settings of the QWP and HWP, which realize the same  measurement in the ideal scenario and  compensate certain first-order errors if the wave plates are not perfect. In particular, we shall design an \emph{error compensation measurement} (ECM) that is capable of  correcting the phase errors of the  QWP and HWP as well as the angle error of the HWP to the second order. The chromatic dispersion is also corrected as a consequence. The ECM is very easy to implement in experiments; no  wave plate with delicate design is necessary. As an application, we apply the ECM to experiments on quantum state tomography on polarization qubits. Experimental results show that the ECM can reduce the systematic error in the estimator by 20 times.

The rest of the paper is organized as follows. In Sec.~\ref{sec:ECM}, we introduce the  ECM to correct the phase errors of the HWP and QWP as well as  the angle error of the HWP. In  Sec.~\ref{sec:QSTapp}, we apply the ECM to experiments on quantum state tomography on polarization qubits and discuss the dependence of the systematic error of the estimator on the imperfections of the wave plates. Section~\ref{sec:sum} summarizes this paper.

\section{\label{sec:ECM}Error compensation measurements}

\begin{figure}
\center{\includegraphics[scale=0.7]{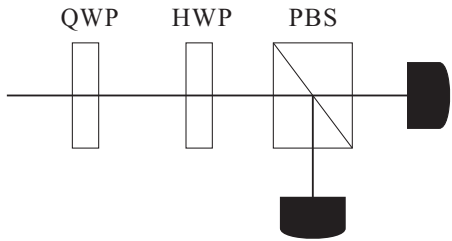}}% Here is how to import EPS art
\caption{\label{configuration: WP-WP} QWP-HWP setting. The measurement device composed of a QWP, an HWP, and a PBS is capable of  realizing arbitrary projective measurements on a polarization qubit.}
\end{figure}

In the case of a qubit, any projective measurement has two outcomes, which can be identified with the two eigenstates of a Pauli operator. A generic Pauli operator has the form $\vect{r}\cdot{\sigma}$, where  $\vect{\sigma}=(\sigma_x,\sigma_y,\sigma_z)^T$
is the vector composed of Pauli matrices
 \begin{equation*}
 \sigma_x=\left(
  \begin{array}{cc}
 0 & 1 \\
 1 & 0 \\
\end{array}
\right),\quad
\sigma_y=\left(
\begin{array}{cc}
 0 & -i \\
 i & 0 \\
 \end{array}
\right),\quad
\sigma_z=\left(
 \begin{array}{cc}
1 & 0 \\
 0 & -1 \\
\end{array}
 \right),
\end{equation*}
and $\vect{r}$ is a unit vector, referred to as the \emph{measurement vector} henceforth.
 The measurement vector $\vect{r}$ is also the Bloch vector of the eigenstate of $\vect{r}\cdot{\sigma}$ with eigenvalue~1. For the polarization qubit, the horizontal and vertical polarization states  $|H\rangle$ and  $|V\rangle$
can be identified with
the two eigenstates of $\sigma_z$ with eigenvalues $1$ and $-1$. In optical experiments, the projective  measurement of $\sigma_z$ can be realized by a PBS.
A generic projective   measurement can be realized by the QWP-HWP setting shown in  Fig.~\ref{configuration: WP-WP}.
Let $q$ and $h$ be the rotation angles  of the   optic axes of the QWP and HWP from the horizontal direction. Then the QWP-HWP setting transforms the state $|H\rangle$ into the state
\begin{equation}\label{psip:QWP-HWP}
\ket{\psi}=\left(
\begin{array}{c}
\cos q\cos t-i\sin q\sin t \\
\sin q\cos t+i\cos q\sin t \\
\end{array}
\right),
\end{equation}
whose Bloch vector is given by
\begin{equation}\label{rp:QWP-HWP}
\vect{r}_0(q,h)=\left(\sin 2q\cos 2t,\sin 2t,\cos 2q\cos 2t\right)^T,
\end{equation}
where $t=2h-q$; see the appendix for more details.

%\begin{equation}\label{deriveative of r:QWP-HWP}
%{\vect{r}(q,h)}=\vect{r}_0(q,h)+\sum_\xi \frac{\partial\vect{r}(q,h)}{\partial\epsilon_\xi}\epsilon_\xi+\sum_{\xi_1,\xi_2} O(\epsilon_{\xi_1}\epsilon_{\xi_2}).
%\end{equation}

In the above analysis, we have assumed that the QWP and HWP are perfect. In real experiments, however, this is not the case. The phase shifts $\delta_q$ and $\delta_h$  of the QWP and HWP may deviate from the ideal values of $90^\circ$ and $180^\circ$. Similarly, the rotation angles of the two wave plates may deviate  from prescribed values.
 Denote by $\epsilon_\xi$ the deviation of $\xi$ from the ideal value, where $\xi$ is one of the four parameters $q, h, \delta_q, \delta_h$. Then the  measurement vector actually realized is
 \begin{equation}\label{eq:rqhep}
 r(q,h,\epsilon):=\vect{r}(q+\epsilon_q, \frac{\pi}{2}+\epsilon_{\delta_q}; h+\epsilon_h, \pi+\epsilon_{\delta_h}),
 \end{equation}
 where the right hand side  is given by  Eq.~(\ref{r:WP-WP}) in the appendix. Note that $\vect{r}_0(q,h)= \vect{r}(q,h,\epsilon=0)$.
The Taylor expansion of $\vect{r}(q,h,\epsilon)$ reads
 \begin{equation}\label{deriveative of r:QWP-HWP}
{\vect{r}(q,h,\epsilon)}=\vect{r}_0(q,h)+\sum_\xi \frac{\partial\vect{r}(q,h,\epsilon)}{\partial{\epsilon_\xi}}\epsilon_\xi+O(\epsilon_\xi^2).
\end{equation}
Here and in the rest of the paper we take the convention that the derivative with respect to $\epsilon_\xi$ is taken at $\epsilon=0$.
The systematic error in the measurement vector may affect the accuracy and reliability of the measurement results. To alleviate these problems, it is crucial to control and reduce such error.

Since device characteristics are not easy to modify in a given experiment, to reduce the systematic error induced by device errors, we need a clever error-compensation design. A starting point of our design is the assumption  that the deviations $\epsilon_{\delta_q}, \epsilon_{\delta_h}, \epsilon_q, \epsilon_h$ do not depend on the rotation angles $q$ and $h$.  This assumption is  reasonable and easy to understand for the phase deviations $\epsilon_{\delta_q}, \epsilon_{\delta_h}$. For the angle deviations $\epsilon_q, \epsilon_h$, it is also a good approximation because
the rotation stage of each wave plate usually can achieve  a much higher precision, say $0.01^\circ$, than the precision, say $0.1^\circ$,  in the calibration angle of the wave plate, so $\epsilon_q, \epsilon_h$ are mainly determined by the calibration process at the beginning of the experiment.

The basic idea of our ECM design can be summarized as follows. Choose  $n$ different angle settings $(q_i,h_i)$ that realize the same measurement in the ideal scenario, that is
\begin{equation}\label{construction: general}
\vect{r}_0(q_1,h_1)=\vect{r}_0(q_2,h_2)=\cdots=\vect{r}_0(q_n,h_n).
\end{equation}
Divide the photons into $n$ groups of equal size and perform the measurement determined by the angle setting $(q_i,h_i)$
on photons in group  $i$. In this way the effective measurement vector realized is the average of $\vect{r}(q_i,h_i,\epsilon)$, that is,
\begin{align}\label{r:error-compensation}
\vect{r}_e&=\frac{1}{n}\sum_{i=1}^n\vect{r}(q_i,h_i,\epsilon) \nonumber \\
&=\vect{r}_0(q_1,h_1)+\frac{1}{n}\sum_\xi\sum_{i=1}^n \frac{\partial\vect{r}(q_i,h_i,\epsilon)}{\partial\epsilon_\xi}\epsilon_\xi+O(\epsilon_\xi^2).
\end{align}
With  a suitable choice of the angle settings $(q_i,h_i)$, the vector $\vect{r}_e$ can be made less sensitive than each $\vect{r}(q_i,h_i,\epsilon)$ to the imperfections of the wave plates. In particular, the first-order error in $\epsilon_\xi$ can be eliminated if the angle settings $(q_i,h_i)$ satisfy the equation
\begin{equation}\label{construction: general2}
\sum_{i=1}^n\frac{\partial\vect{r}(q_i,h_i,\epsilon)}{\partial\epsilon_\xi}=0.
\end{equation}

In the rest of this section, we present an ECM design that is capable of correcting the first-order errors in $\epsilon_{\delta_q}, \epsilon_{\delta_h}, \epsilon_h$ simultaneously. To start with,  ECM designs for correcting these errors separately are introduced.

\subsection{ECM design for $\epsilon_h$ }
The partial derivative of $\vect{r}(q,h,\epsilon)$ with respect to $\epsilon_h$ (at $\epsilon=0$) is
\begin{align}\label{derivative of r:h}
&\frac{\partial\vect{r}(q,h,\epsilon)}{\partial\epsilon_h}=\frac{\partial\vect{r}_0(q,h)}{\partial h}\nonumber\\
&=4\left(-\sin 2q\sin 2t,\cos 2t,-\cos 2q\sin 2t\right)^T.
\end{align}
where we have applied   Eq.~(\ref{rp:QWP-HWP}).
To compensate the first-order error in  $\epsilon_h$, we need two different
angle settings. Substituting Eqs.~ (\ref{rp:QWP-HWP}) and (\ref{derivative of r:h}) into the two error compensation conditions in Eqs.~(\ref{construction: general}) and (\ref{construction: general2}) with  $n=2$, we get
\begin{equation}\label{construction: h}
\begin{aligned}
\sin 2q_1\cos 2t_1&=\sin 2q_2\cos 2t_2,\\
\sin 2t_1&=\sin 2t_2,\\
\cos 2q_1\cos 2t_1&=\cos 2q_2\cos 2t_2,\\
-4\sin 2q_1\sin 2t_1&=4\sin 2q_2\sin 2t_2,\\
4\cos 2t_1&=-4\cos 2t_2,\\
-4\cos 2q_1\sin 2t_1&=4\cos 2q_2\sin 2t_2.
\end{aligned}
\end{equation}
The set of equations can be simplified as
\begin{equation}\label{construction: h independent}
\begin{aligned}
\sin 2t_1&=\sin 2t_2,\\
\cos 2t_1&=-\cos 2t_2,\\
\sin 2q_1&=-\sin 2q_2,\\
\cos 2q_1&=-\cos 2q_2,\\
\end{aligned}
\end{equation}
with solutions given by
\begin{equation}\label{solution:h}
\begin{aligned}
q_2&=q_1+(k+0.5)\pi,\quad
h_2=q_1-h_1+\frac{k^\prime}{2}\pi,
\end{aligned}
\end{equation}
 where $k$ and $k^\prime$ are arbitrary integers.

\subsection{ECM design for $\epsilon_{\delta_q}$}
The partial derivative of $\vect{r}(q,h,\epsilon)$ with respect to
 $\epsilon_{\delta_q}$ can be computed according to  Eq.~(\ref{eq:rqhep}) as well as  Eqs.~(\ref{r:WP-WP}) and (\ref{A:WP-WP}) in the appendix, with the result
\begin{equation}\label{deriveative of r:delta_1}
\frac{\partial\vect{r}(q,h,\epsilon)}{\partial \epsilon_{\delta_q}}=\left(-\cos 2q\sin 2t,0,\sin 2q\sin 2t\right)^T.
\end{equation}
Again two
angle settings are sufficient for compensating the first-order error in $\epsilon_{\delta_q}$. Substituting Eqs.~ (\ref{rp:QWP-HWP}) and (\ref{deriveative of r:delta_1}) into the two error compensation conditions in Eqs.~(\ref{construction: general}) and (\ref{construction: general2}) yields
\begin{equation}\label{construction: delta1}
\begin{aligned}
\sin 2q_1\cos 2t_1&=\sin 2q_2\cos 2t_2,\\
\sin 2t_1&=\sin 2t_2,\\
\cos 2q_1\cos 2t_1&=\cos 2q_2\cos 2t_2,\\
-\cos 2q_1\sin 2t_1&=\cos 2q_2\sin 2t_2,\\
\sin 2q_1\sin 2t_1&=-\sin 2q_2\sin 2t_2.
\end{aligned}
\end{equation}
Interestingly,  Eq.~(\ref{construction: delta1}) is equivalent to  Eq.~(\ref{construction: h independent}) and thus  has the same solution  as in Eq.~(\ref{solution:h}).

\subsection{ECM design for $\epsilon_{\delta_h}$}
The partial derivative of $\vect{r}(q,h,\epsilon)$ with respect to
 $\epsilon_{\delta_h}$ can be computed according to  Eq.~(\ref{eq:rqhep}) as well as  Eqs.~(\ref{r:WP-WP}) and (\ref{A:WP-WP}) in the appendix, with the result
\begin{equation}\label{deriveative of r:delta_2}
\frac{\partial\vect{r}(q,h,\epsilon)}{\partial \epsilon_{\delta_h}}=\left(-\cos 2q\sin 2h,0,\sin 2q\sin 2h\right)^T.
\end{equation}
The  error-compensation conditions can be derived
following the same reasoning  as in the previous two sections, with the result
\begin{equation}\label{construction: delta2 independent}
\begin{aligned}
\text{tan}2q_1&=\text{tan}2q_2,\\
\sin 2q_1\cos 2t_1&=\sin 2q_2\cos 2t_2,\\
\sin 2t_1&=\sin 2t_2,\\
\sin 2q_1\sin 2h_1&=-\sin 2q_2\sin 2h_2.
\end{aligned}
\end{equation}
The solutions are given by
\begin{equation}\label{solution:delta_h}
\begin{aligned}
q_2&=q_1+k\pi,\quad
h_2=h_1+(k^\prime+0.5)\pi,
\end{aligned}
\end{equation}
where $k$ and $k^\prime$ are arbitrary integers.

\subsection{ECM design for $\epsilon_q$}

The partial derivative of $\vect{r}(q,h,\epsilon)$ with respect to $\epsilon_q$ is
\begin{align}\label{deriveative of r:q}
&\frac{\partial\vect{r}(q,h,\epsilon)}{\partial\epsilon_q}=\frac{\partial\vect{r}_0(q,h)}{\partial q}\nonumber\\
&=2\left(\cos (4h-4q),-\cos 2t,\sin (4h-4q)\right)^T.
\end{align}
The error-compensation conditions can be expressed as follows,
\begin{equation}\label{construction: q independent}
\begin{aligned}
\sin 2t_1&=\sin 2t_2,\\
\cos 2t_1&=-\cos 2t_2,\\
\sin 2q_1&=-\sin 2q_2,\\
\cos 2q_1&=-\cos 2q_2,\\
\cos (4h_1-4q_1)&=-\cos (4h_2-4q_2),\\
\sin (4h_1-4q_1)&=-\sin (4h_2-4q_2).
\end{aligned}
\end{equation}
Unfortunately, this set of equations cannot be satisfied simultaneously except when  $q_1$ and $h_1$ take on certain special values.

\begin{figure}
\center{\includegraphics[scale=0.51]{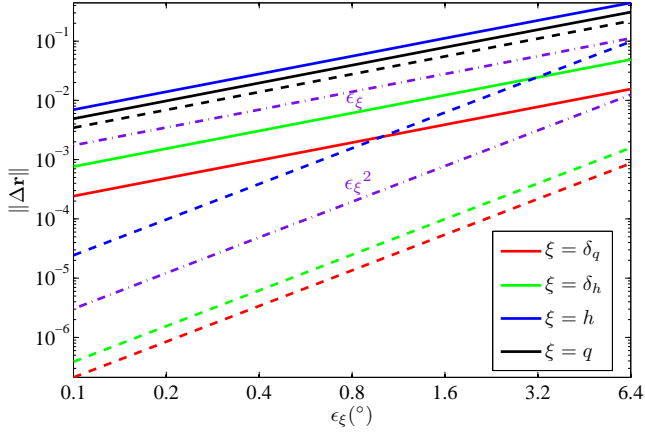}}% Here is how to import EPS art
\caption{\label{error_of_r}(color online) Systematic errors versus device errors in NCM and ECM.
The systematic errors in the measurement vector $\vect{r}_0(q_1=30^\circ,h_1=13^\circ)$ in the NCM scheme (solid line) and ECM scheme (dashed line) are plotted as functions of  deviations  in  $\delta_q$ (red), $\delta_h$ (green), $h$ (blue), and $q$ (black). The linear function $\epsilon_\xi$ and the quadratic function $\epsilon_\xi^2$ are plotted in purple dash-dotted lines to reflect the scaling behaviors of the systematic errors. In the NCM scheme, the systematic errors are linear in the deviations. In the ECM scheme, by contrast, they are quadratic in  the deviations except for $\epsilon_q$.
}
\end{figure}

%
%
%The two purple dash-dot lines are plotted as the first and second order of $\epsilon_\xi$ for the identification of the order of the systematic error. In the NCM scheme (solid line), the systematic error with respect to the four device errors is parallel to $\epsilon_\xi$ and is $O(\epsilon_\xi)$. In the ECM scheme, $||\Delta\vect{r}||=O({\epsilon_\xi}^2)$ except for $\xi=q$. Even though the systematic error in the ECM scheme is $O(\epsilon_q)$, ECM still outperforms NCM.

\subsection{ECM design for $\epsilon_h$, $\epsilon_{\delta_q}$, and $\epsilon_{\delta_h}$ simultaneously}

Based on the separate ECM designs for $\epsilon_h$, $\epsilon_{\delta_q}$, and $\epsilon_{\delta_h}$, we can now introduce an ECM design that is capable of correcting the first-order errors in  $\epsilon_h$, $\epsilon_{\delta_q}$, and $\epsilon_{\delta_h}$ simultaneously.
The design is specified  by the following angle settings,
\begin{equation}\label{EQ: series to compensate all}
\begin{aligned}
q_2&=q_1+\frac{1}{2}\pi,& h_2& =q_1-h_1;\\
q_3&=q_1,&  h_3&=h_1+\frac{1}{2}\pi;\\
q_4&=q_2,&\quad h_4&=h_2+\frac{1}{2}\pi.
\end{aligned}
\end{equation}
To verify our claim, note that the two angle settings $(q_3,h_3)$ and $(q_1,h_1)$ compensate the first-order error in  $\epsilon_{\delta_h}$ according to Eq.~(\ref{solution:delta_h}), and  so do the two angle settings $(q_4,h_4)$ and $(q_2,h_2)$. Meanwhile, the two angle settings $(q_2,h_2)$ and $(q_1,h_1)$ compensate the first-order errors in  $\epsilon_h$ and $\epsilon_{\delta_q}$, and so do the two angle settings $(q_4,h_4)$ and $(q_3,h_3)$. Therefore, the four settings can correct the first-order errors in  $\epsilon_{h},\epsilon_{\delta_q}, \epsilon_{\delta_h}$ simultaneously, as illustrated  in Fig.~\ref{error_of_r}. Incidentally, our ECM design also corrects the first-order chromatic dispersion, which affects the phase shifts $\delta_q$ and  $\delta_h$ of the QWP and HWP.

%Note that as
%\begin{equation*}
%\frac{\partial\vect{r}_e}{\partial \lambda}=\frac{\partial\vect{r}_e}{\partial \epsilon_{\delta_q}}\frac{\partial\epsilon_{\delta_q}}{\partial \lambda}+\frac{\partial\vect{r}_e}{\partial \epsilon_{\delta_h}}\frac{\partial\epsilon_{\delta_h}}{\partial \lambda}=0,
%\end{equation*}
%where $\lambda$ is the wave length, ECM also compensates the fist order of chromatic dispersion.

Our ECM design also reduces the error induced by  $\epsilon_q$ although it cannot correct the error. To see this,
we can compare the norm of the partial derivative $\frac{\partial\vect{r}_e}{\partial\epsilon_q}$ with that of $\frac{\partial\vect{r}}{\partial\epsilon_q}$.
Note that
\begin{align}
\frac{\partial\vect{r}_e}{\partial\epsilon_q}&=\frac{1}{4}\sum_{i=1}^4\frac{\partial\vect{r}(q_i,h_i,\epsilon)}{\partial\epsilon_q} \nonumber \\
&=2\left(\cos 2q_1\cos 2t_1,0,-\sin 2q_1\cos 2t_1\right)^T \nonumber     \\
&=2U_0\vect{r}_0(q_1,h_1),
\end{align}
 where
\begin{equation}\label{EQ: U0}
U_0=\left(
\begin{array}{ccc}
0 & 0 & 1 \\
0 & 0 & 0 \\
-1 & 0 & 0 \\
\end{array}
\right).
\end{equation}
We have
\begin{equation}
\left\|\frac{\partial\vect{r}_e}{\partial\epsilon_q}\right\|^2=4\cos^22t_1=4(1-r_y^2).
\end{equation}
This value is at least two times smaller than the value without error compensation \begin{equation}
\left\|\frac{\partial\vect{r}(q_1,h_1,\epsilon)}{\partial\epsilon_q}\right\|^2=4+4\cos^22t_1=4(2-r_y^2),
\end{equation}
which follows  from Eq.~(\ref{deriveative of r:q}); see  Fig.~\ref{error_of_r}. To appreciate the improvement of the ECM over the NCM, let us consider   a complete set of mutually unbiased measurements. The corresponding measurement vectors $\vect{r}^{(i)}$ for  $i=1,2,3$ form an orthonormal basis, which implies that $\sum_i\bigl(r_y^{(i)}\bigr)^2=1$.  Therefore,  $\sum_i\|\frac{\partial\vect{r}_e^{(i)}}{\partial\epsilon_q}\|^2=\sum_i 4\bigr[1-\bigl(r_y^{(i)}\bigr)^2\bigr]=8$, which is only two fifths  of the original value $\sum_i\|\frac{\partial\vect{r}^{(i)}}{\partial\epsilon_q}\|^2=20$.

\section{\label{sec:QSTapp}Applications in quantum state tomography}
\subsection{Systematic error in quantum state tomography}
Quantum state tomography is a procedure for  inferring  the state of a  quantum system from  quantum measurements and data processing \cite{Pari04quantum,Lvov09continuous,Zhu12quantum}.  A measurement is informationally complete if all states of the given quantum system can be distinguished by probabilities of measurement outcomes or, equivalently, if the measurement outcomes span the operator space.
When a large number of copies of the unknown quantum state are measured, the frequencies of measurement outcomes converge to the corresponding probabilities. The quantum state can be reconstructed reliably if the measurement is informationally complete and the measurement devices are perfect. In practice, however, the measurement devices are never perfect, and the state reconstructed may deviate from the true state even if arbitrarily large number of copies of the state are available for measurements. To achieve reliable quantum state tomography,  therefore, it is indispensable to understand and reduce the systematic error induced by device errors. Here we show that the ECM design introduced in the previous section can be applied to reduce the systematic error in quantum state tomography on polarization qubits. To highlight the main point, we neglect the statistical error in the following analysis.

Suppose the state of the qubit is characterized by the density operator
$\rho=(1+\vect{s}\cdot\vect{\sigma})/2$, where $\vect{s}$ is the Bloch vector. To reconstruct the state $\rho$ is equivalent to reconstruct the Bloch vector $\vect{s}$. To achieve this task, we can perform three projective measurements corresponding to
 three Pauli operators $\vect{r}_0^{(i)}\cdot\vect{\sigma}$ for $i=1,2,3$, where  the measurement vectors $\vect{r}_0^{(i)}$ are  linearly independent. In the ideal scenario, the measurement of $\vect{r}_0^{(i)}\cdot\vect{\sigma}$ determines the projection of the Bloch vector $\vect{s}$ on the direction $\vect{r}_0^{(i)}$,
\begin{equation}
{m}_0^{(i)}=\vect{r}_0^{(i)}\cdot\vect{s}.
\end{equation}
The Bloch vector $\vect{s}$ can be reconstructed via the formula
\begin{equation}
\hat{\vect{s}}=({R_0}^T)^{-1}\vect{m}_0=\vect{s},
\end{equation}
where
\begin{equation}\label{definition of R_0}
\begin{aligned}
\quad R_0=(\vect{r}_0^{(1)},\vect{r}_0^{(2)}, \vect{r}_0^{(3)}), \quad\vect{m}_0=(m_0^{(1)},m_0^{(2)},m_0^{(3)})^T.
\end{aligned}
\end{equation}
Due to device errors,  the result we really get is
\begin{equation}\label{expectation value of r}
{m}^{(i)}=\vect{r}^{(i)}\cdot\vect{s},
\end{equation}
where $\vect{r}^{(i)}$ is the measurement vector actually realized.
Consequently, the estimator
\begin{equation}\label{estimation of s}
\hat{\vect{s}}=({R_0}^T)^{-1}\vect{m}=\vect{s}+({R_0}^T)^{-1}(\vect{m}-\vect{m}_0)
\end{equation}
has a systematic error,
\begin{align}\label{systematic error of s}
\Delta\vect{s}&=\hat{\vect{s}}-\vect{s}=({R_0}^T)^{-1}(R-R_0)^T\vect{s}\nonumber\\
&=\sum_\xi({R_0}^T)^{-1}\frac{\partial R^T}{\partial\epsilon_\xi}\vect{s}\epsilon_\xi+O(\epsilon_\xi^2),
\end{align}
where $R=(\vect{r}^{(1)},\vect{r}^{(2)}, \vect{r}^{(3)})$.
This equation  clearly shows how the systematic error of the estimator $\hat{\vect{s}}$ depends on the measurement bases, the device errors, and the quantum state. In the NCM scheme, $R-R_0$ is linear in the deviations $\epsilon_q, \epsilon_h, \epsilon_{\delta_q}, \epsilon_{\delta_h}$, so $\|\Delta\vect{s}\|^2$ is quadratic in these deviations.

To analyze the systematic error in the  ECM scheme, let $R_e=(\vect{r}_e^{(1)},\vect{r}_e^{(2)}, \vect{r}_e^{(3)})$. Noticing that  $\frac{\partial R_e}{\partial\epsilon_\xi}=0$ except when $\xi=q$, we deduce that
	\begin{equation}\label{EQ: error s ECM}
	\begin{aligned}
	\Delta\vect{s}&=({R_0}^T)^{-1}(R_e-R_0)^T\vect{s}\\
	&=({R_0}^T)^{-1}\frac{\partial {R_e}^T}{\partial\epsilon_q}\vect{s}\epsilon_q+O(\epsilon_\xi^2)\\
	&=2({R_0}^T)^{-1}(U_0R_0)^T\vect{s}\epsilon_q+O(\epsilon_\xi^2)\\
	&=2{U_0}^T\vect{s}\epsilon_q+O(\epsilon_\xi^2),
	\end{aligned}
	\end{equation}
	where $U_0$ is defined in  \eref{EQ: U0}.  It follows that
	\begin{equation}\label{systematic error of state in error-compensation}
	\|\Delta\vect{s}\|^2=4({s_x}^2+{s_z}^2){\epsilon_q}^2+O(\epsilon_{q}\epsilon_\xi^2).
	\end{equation}
The systematic error $\|\Delta\vect{s}\|^2$	 for the ECM scheme is quadratic only in $\epsilon_q$, as illustrated in the upper plot of Fig.~\ref{s1}.
Interestingly, the quadratic part of  $\|\Delta\vect{s}\|^2$
is independent of  measurement bases  and depends only on the Bloch vector  $\vect{s}$ of the quantum state and the deviation angle  $\epsilon_q$.
When the  Bloch vector is aligned along the $y$ direction (i.e., $\vect{s}=(0,s_y,0)^T$),  $\Delta\vect{s}$ is quadratic in  $\epsilon_q, \epsilon_h, \epsilon_{\delta_q}, \epsilon_{\delta_h}$, so $\|\Delta\vect{s}\|^2$ is quartic  in these parameters,   as illustrated  in the lower plot of Fig.~\ref{s1}.

\subsection{Experimental results and discussion}

%\begin{figure}
%  \centering
%
%\center{\includegraphics[scale=0.545]{errorS1}}
%  \caption[]
%  {\label{s1}(color online) Systematic error in quantum state tomography of $\vect{s}_1$. The systematic error in estimating $\vect{s}_1$ is calculated with respect to imperfections of $\delta_q$ (red), $\delta_h$ (green), $h$ (blue) and $q$ (black) in the NCM scheme (solid line) and ECM scheme (dashed line) separately. Diamonds and circles are the averages of experimental results from 5 runs with NCM and ECM schemes; error bars are the standard deviation, some of which are not visible because they are too small. Systematic error in NCM are all $O({\epsilon_{\xi}}^2)$ and in ECM are $O({\epsilon_{\xi}}^4)$ other than $\xi=q$. Systematic error below $8\times10^{-6}$ (gray region) is dominated by statistical fluctuation.}
%\end{figure}
%
%
%\begin{figure}
%  \centering
%
%\center{\includegraphics[scale=0.5]{errorS2}}
%  \caption[]
%  {\label{s2}(color online) Systematic error in quantum state tomography of $\vect{s}_2$. Different from $\vect{s}_1$, the systematic error in the estimation of $\vect{s}_2$  with respect to $\xi=q$ in ECM is $O({\epsilon_{\xi}}^4)$ rather than $O({\epsilon_{\xi}}^2)$ due to the special form of $\vect{s}_2$. Diamonds and circles are the mean of five experimental trials in NCM and ECM schemes; error bars are the standard deviation, some of which are not visible because they are too small. Systematic error below $8\times10^{-6}$ (gray region) is dominated by statistical fluctuation.}
%\end{figure}

\begin{figure}
  \centering
\center{\includegraphics[scale=0.545]{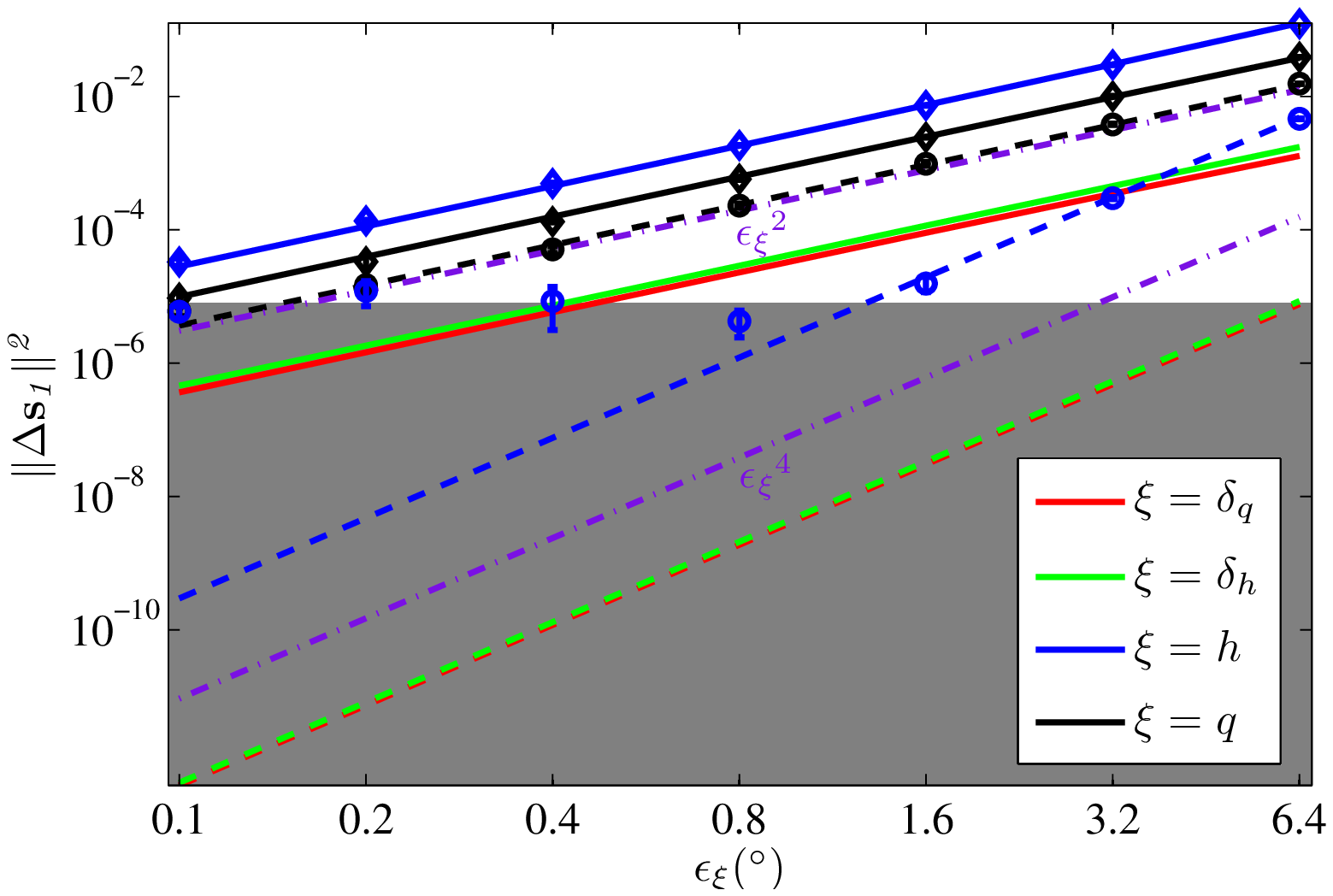}}
\center{\includegraphics[scale=0.5]{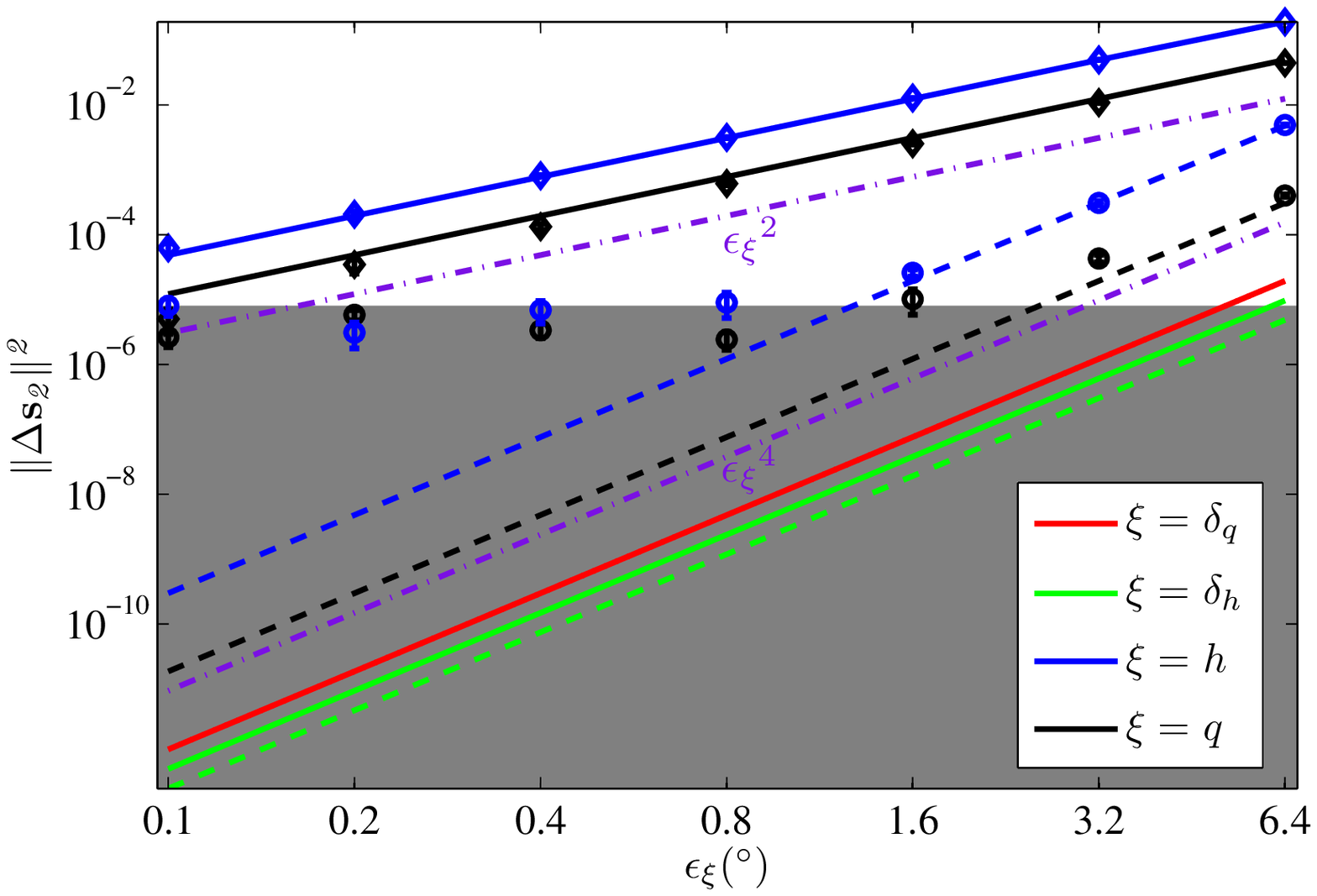}}
  \caption[]
  {\label{s1}(color online) Systematic errors in quantum state tomography of  polarization qubits.
The systematic errors  of the estimators  in the NCM scheme (solid line) and ECM scheme (dashed line) are plotted as functions of  deviations  in  $\delta_q$ (red), $\delta_h$ (green), $h$ (blue), and $q$ (black).
Diamonds and circles show the averages of experimental results in five runs with NCM and ECM schemes, respectively; the error bars denote the standard deviations, some of which are not visible because the standard deviations are too small.  Systematic error below $8\times10^{-6}$ (gray region) is dominated by statistical fluctuation. In the upper plot, the state has Bloch vector $\vect{s}_1=(0.346,-0.446,0.425)^T$. The systematic error $\|\Delta\vect{s}_1\|^2$ in the NCM scheme is quadratic in the deviations. In the ECM scheme, by contrast, it is quadratic in $\epsilon_q$ but   quartic in other three deviations.
In the lower plot, the state has Bloch vector $\vect{s}_2=(0,0.707,0)^T$. Due to this  special choice, the systematic error $\|\Delta\vect{s}_1\|^2$ in the NCM scheme is quadratic in $\epsilon_h, \epsilon_q$ but quartic in $\epsilon_{\delta_h}, \epsilon_{\delta_q}$.
In the ECM scheme, $\|\Delta\vect{s}_1\|^2$ is quartic in all the four deviations. }
\end{figure}

In this section we present some experimental results on quantum state tomography on polarization qubits to demonstrate the efficacy of our ECM design in reducing the systematic error.
The experimental setup is the same as that in Ref.~\cite{Hou15experimental}, as illustrated in Fig.~2 therein. A pair of H-polarized photons with wave length $\lambda=808$~nm is created via spontaneous parametric down-conversion (SPDC) process. One photon is detected by a single-photon detector and heralds the generation of the other photon. The polarization state of the other photon is prepared by a combination of a half-wave plate, a quartz crystal, and a quarter-wave plate.
Projective measurements on the photon are realized  by the QWP-HWP setting illustrated in Fig.~\ref{configuration: WP-WP}. Since the phase errors of the QWP and HWP in the QWP-HWP setting are fixed in a given experiment, we can only test the relationship between the systematic error in quantum state tomography and the angle errors of the QWP and HWP by intentionally adding  errors to the initially well-calibrated optic axes.

In the experiments, we prepared two polarization states with Bloch vectors  $\vect{s}_1=(0.346,-0.446,0.425)^T$ and $\vect{s}_2=(0,0.707,0)^T$  by setting  the rotation angle of the QWP in the  state-preparation module at $19.57^\circ$ and  $-45^\circ$, respectively  (see Fig.~2 in Ref.~\cite{Hou15experimental}). The Bloch vectors of the two states have the same length of $0.707$.
The measurement scheme consists of projective measurements on the three Pauli operators $\sigma_x, \sigma_y, \sigma_z$, realized with the angle settings $q^{(1)}=45^\circ$, $ h^{(1)}=22.5^\circ$; $q^{(2)}=0^\circ$, $h^{(2)}=22.5^\circ$; and $q^{(3)}=0^\circ$, $h^{(3)}=0^\circ$, respectively.  The corresponding  measurement vectors are  given by  $\vect{r}_0^{(1)}=(1,0,0)^T$, $\vect{r}_0^{(2)}=(0,1,0)^T$, and $\vect{r}_0^{(3)}=(0,0,1)^T$, so that $R_0=I_{3\times3}$. To reduce the statistical fluctuation, $3\times10^6$ photons
are measured in each setting and the same measurements are repeated  five times.

To analyze the systematic error, we need to compute the partial derivatives of $R$ with respect to  $h, q, ,\delta_h, \delta_q$.
According to
\esref{derivative of r:h}, (\ref{deriveative of r:delta_1}), (\ref{deriveative of r:delta_2}), and (\ref{deriveative of r:q}), we have
\begin{align*}
\frac{\partial R}{\partial\epsilon_h}&=\left(
                                                    \begin{array}{ccc}
                                                      0 & 0 & 0 \\
                                                      4 & 0 & 4 \\
                                                      0 & -4 & 0 \\
                                                    \end{array}
                                                  \right),&
\frac{\partial R}{\partial \epsilon_{\delta_q}}&=\left(
                                       \begin{array}{ccc}
                                         0 & -1 & 0 \\
                                         0 & 0 & 0 \\
                                         0 & 0 & 0 \\
                                       \end{array}
                                     \right),\\
\frac{\partial R}{\partial \epsilon_{\delta_h}}&=\frac{1}{\sqrt{2}}\left(
                                       \begin{array}{ccc}
                                         0 & -1 & 0 \\
                                         0 & 0 & 0 \\
                                         1 & 0 & 0 \\
                                       \end{array}
                                     \right),&
                                     \frac{\partial R}{\partial\epsilon_q}&=\left(
                                       \begin{array}{ccc}
                                         0 & 0 & 2 \\
                                         -2 & 0 & -2 \\
                                         -2 & 2 & 0 \\
                                       \end{array}
                                     \right).
\end{align*}
The quadratic part of the systematic error $\|\Delta\vect{s}\|^2$ in the NCM scheme can be computed according to
these equations and Eq.~(\ref{systematic error of s}), with the result
\begin{align}
\|\Delta\vect{s}\|^2&\approx(32s_y^2+16s_z^2)\epsilon_h^2+\frac{1}{2}(s_x^2+s_z^2)\epsilon_{\delta_h}^2+s_x^2\epsilon_{\delta_q}^2\nonumber\\
&\quad +4(s_x^2+2s_y^2+2s_z^2+2s_ys_z-2s_xs_y)\epsilon_q^2,
\end{align}
where we have assumed that the errors induced by $\epsilon_h, \epsilon_q, \epsilon_{\delta_h}, \epsilon_{\delta_q}$ are independent.
The systematic error
 $\|\Delta\vect{s}_1\|^2$ is equal to  $9.2{\epsilon_h}^2+3.2{\epsilon_q}^2+0.12{\epsilon_{\delta_q}}^2+0.15{\epsilon_{\delta_h}}^2$ up to the second order (see Fig.~\ref{s1}). With  typical angle errors of $0.1^{\circ}$ and phase errors of $1.2^{\circ}$ for both the QWP and HWP, the second-order  errors due to $\epsilon_h,\epsilon_q,\epsilon_{\delta_q}$ and $\epsilon_{\delta_h}$ are  $2.8\times10^{-5}$, $9.8\times10^{-6}$, $5.3\times10^{-5}$, and $6.6\times10^{-5}$, respectively, which sum up to $1.6\times10^{-4}$.
  The systematic error
 $\|\Delta\vect{s}_2\|^2$ is equal to  $16{\epsilon_h}^2+4{\epsilon_q}^2$ up to the second order.
 The errors due to $\epsilon_h$ and $\epsilon_q$ are  $4.9\times10^{-5}$ and $1.2\times10^{-5}$, respectively,  which sum  up to $6.1\times10^{-5}$.

 By contrast, the systematic error in the ECM scheme is given by Eq.~(\ref{systematic error of state in error-compensation}).
 The systematic error  $\|\Delta\vect{s}_1\|^2$ reduces to  $1.2{\epsilon_q}^2$, which equals $3.7\times10^{-6}$ for $\epsilon_q=0.1^{\circ}$.  Due to the special choice of $\vect{s}_2$, the quadratic part of $\|\Delta\vect{s}_2\|^2$
 vanishes (see Fig.~\ref{s1}).

Fig.~\ref{s1}  shows the experimental result on the systematic errors associated with  the NCM and ECM schemes.
Due to  the statistical error associated with finite number of measurements ($3\times10^{6}$ in our experiments),
 the systematic error of the ECM scheme is limited to $8\times10^{-6}$ instead of $4\times10^{-6}$ and  $10^{-10}$ for the two states predicted theoretically. Nevertheless, the ECM scheme  successfully corrects  dominant errors induced by
 $\epsilon_{\delta_q}$, $\epsilon_{\delta_h}$, and $\epsilon_h$, thereby
 achieving 20-fold improvement over the NCM scheme.

% In the NCM scheme, the systematic error
% $\|\Delta\vect{s}_1\|^2$ is equal to  $9.2{\epsilon_h}^2+3.2{\epsilon_q}^2+0.12{\epsilon_{\delta_q}}^2+0.15{\epsilon_{\delta_h}}^2$ up to the second order (see Fig.~\ref{s1}). With  typical angle errors of $0.1^{\circ}$ and phase errors of $1.2^{\circ}$ for both the QWP and HWP, the second-order  errors due to $\epsilon_h,\epsilon_q,\epsilon_{\delta_q}$ and $\epsilon_{\delta_h}$ are  $2.8\times10^{-5}$, $9.8\times10^{-6}$, $5.3\times10^{-5}$, and $6.6\times10^{-5}$, respectively, which sum up to $1.6\times10^{-4}$.
% In the ECM scheme, according to Eq.~(\ref{systematic error of state in error-compensation}), the systematic error reduces to  $1.2{\epsilon_q}^2$, which equals $3.7\times10^{-6}$ for $\epsilon_q=0.1^{\circ}$ (see Fig.~\ref{s1}).
%% Another state with Bloch vector $\vect{s}_2=(0,0.707,0)^T$ was prepared  by setting  the rotation angle of QWP in the  state-preparation module at $-45^\circ$.
%  The systematic error
% $\|\Delta\vect{s}_2\|^2$ in the NCM scheme is equal to  $16{\epsilon_h}^2+4{\epsilon_q}^2$ up to the second order.
% The errors due to $\epsilon_h$ and $\epsilon_q$ are  $4.9\times10^{-5}$ and $1.2\times10^{-5}$, respectively,  which sum  up to $6.1\times10^{-5}$.  In the ECM scheme, the second-order error vanishes  due to the special choice of $\vect{s}_2$.

\section{\label{sec:sum}Summary}

We have introduced an ECM design on polarization qubits. Our ECM design is able to reduce the systematic error to the second order of the phase errors of the  HWP and QWP as well as  the angle error of the HWP. Although it cannot eliminate the first-order error induced by the angle error of the QWP, it decreases  the error by several times. In typical optical  experiments, phase errors are usually more serious than angle errors. Therefore, our ECM design is quite helpful in reducing the systematic error.
As an application, the ECM design was  employed  in an experiment on quantum state tomography of the polarization qubit, achieving a 20-fold reduction in the systematic error of the estimator. Our study may find applications in optical experiments that demand high precision.

\section*{Acknowledgments}

The work at USTC is supported by National Fundamental Research Program (Grants No. 2011CBA00200 and No. 2011CB9211200), National Natural Science Foundation of China (Grants No. 61108009 and No. 61222504). H.Z. is supported  by Perimeter Institute for Theoretical Physics. Research at Perimeter Institute is supported by the Government of Canada through Industry Canada and by the Province of Ontario through the Ministry of Research and Innovation.

\bibliographystyle{apsrev4-1}
\bibliography{all_references_hou}

\newpage
\onecolumngrid
\numberwithin{equation}{section}
\setcounter{equation}{0}
\appendix

\section{Qubit measurements with two wave plates}\label{two wave plate setting}

In optical experiments, wave plates and polarizing beam splitters are used to realize projective measurements on polarization qubits.
Two wave plates and one PBS are  sufficient to realize arbitrary projective measurements on a single qubit. One such setup is shown in
Fig.~\ref{configuration: WP-WP} (here we do not assume that the two wave plates are a quarter-wave plate (QWP) and a half-wave plate (HWP)). In the absence of the two wave plates, the PBS realizes the projective  measurement on the basis composed of
the horizontal and vertical  polarization states $|H\rangle$ and $|V\rangle$.
In general, let $\delta_q, \delta_h$ be the phase shifts of the left and the right  wave plates, and let  $q, h$ be the rotation  angles of the two wave plates from the horizontal direction. Then the two wave plates transform the state $\ket{H}$ into
\begin{equation}\label{psi:WP-WP}
  \ket{\psi}=U(q,\delta_q)U(h,\delta_h)\ket{H},
\end{equation}
where $U(\theta,\delta)$ is the  unitary transformation  realized by a wave plate with rotation angle $\theta$ and phase shift $\delta$,
\begin{equation}\label{unitary}
  U(\theta,\delta)=\left(
      \begin{array}{cc}
        \cos^2\theta+e^{i\delta}\sin^2\theta & \frac{1}{2}(1-e^{i\delta})\sin 2\theta \\
        \frac{1}{2}(1-e^{i\delta})\sin 2\theta & \sin^2\theta+e^{i\delta}\cos^2\theta \\
      \end{array}
    \right).
\end{equation}
The Bloch vector $\vect{r}$ of  $|\psi\rangle\langle\psi|$ is
\begin{equation}\label{r:WP-WP}
\vect{r}(q,\delta_q;h,\delta_h)=\vect{A}_1\cos \delta_q+\vect{A}_2\sin \delta_q+\vect{A}_3,
\end{equation}
where
\begin{equation}\label{A:WP-WP}
\begin{aligned}
\vect{A}_1&=\frac{1}{4}\left(
                                                                \begin{array}{l}
(\cos  \delta_h-1)[\sin (4q-4h)-\sin 4q-\sin 4h]-2\sin 4q \\
-4\sin \delta_h\sin 2h \\
(\cos \delta_h-1)[\cos (4q-4h)-\cos 4q-\cos 4h+1]-2\cos 4q+2\\
                                                                \end{array}
                                                              \right),\\
\vect{A}_2&=\left(
                                                                \begin{array}{l}
\sin \delta_h\cos 2q\sin 2h\\
-(\cos \delta_h-1)\cos (2q-2h)\sin 2h-\sin 2q \\
-\sin \delta_h\sin 2q\sin 2h\\
                                                                \end{array}
                                                              \right),\\
\vect{A}_3&=\frac{1}{4}\left(
                                                                \begin{array}{l}
-(\cos \delta_h-1)[\sin (4q-4h)-\sin 4q+\sin 4h]+2\sin 4q \\
0 \\
-(\cos \delta_h-1)[\cos (4q-4h)-\cos 4q+\cos 4h-1]+2\cos 4q+2\\
                                                                \end{array}
                                                              \right).
\end{aligned}
\end{equation}

When the left and the right wave plates in Fig.~\ref{configuration: WP-WP} are a QWP and a HWP, that is, $\delta_q=90^\circ$, $\delta_h=180^\circ$, the state $\ket{\psi}$ turns out to be
\begin{equation}\label{psi:QWP-HWP}
  \ket{\psi}=U(q,\frac{\pi}{2})U(h,\pi)\ket{H}=\left(
                                                                \begin{array}{c}
\cos q\cos (q-2h)+i\sin q\sin (q-2h) \\
\sin q\cos (q-2h)-i\cos q\sin (q-2h) \\
                                                                \end{array}
                                                              \right),
\end{equation}
and its Bloch vector is
\begin{equation}\label{r:QWP-HWP}
\vect{r}_0(q,h)=\vect{r}(q,\frac{\pi}{2};h,\pi)=\left(\sin 2q\cos (4h-2q),\sin (4h-2q),\cos 2q\cos (4h-2q)\right)^T.
\end{equation}

\end{document}